COSPAR02-A-00897

# WORLD SPACE OBSERVATORY/ULTRAVIOLET (WSO/UV): PROGRESS REPORT

Willem Wamsteker [1] and Hans J. Haubold [2]

[1] *European Space Agency, Satellite Tracking Station VILSPA, P.O. Box 50727, 28080 Madrid, Spain*
[2] *Office for Outer Space Affairs, United Nations, Vienna International Centre, P.O. Box 500, 1400 Vienna, Austria*

## ABSTRACT

The World Space Observatory/Ultraviolet (WSO/UV) represents a new mission implementation model for large space missions for astrophysics. The process has been brought up to enable, fully scientific needs driven, a logic to be applied to the demands for large collection powers required to undertake space missions which are complementary to the continuously increasing sensitivity of ground-based telescopes. One of the assumptions associated with the idea of a WSO is to avoid the excessive complexity required for multipurpose missions. Although there may exist purely technological or programmatic policy issues, which would suggest such more complex missions to be more attractive, many other aspects, which do not need to be explored in this report, may argue against such a mission model. Following this precept and other reasons, the first implementation model for a WSO has been done for the ultraviolet domain WSO/UV. WSO/UV is a follow-up project of the UN/ESA Workshops on Basic Space Science, organised annually since 1991.

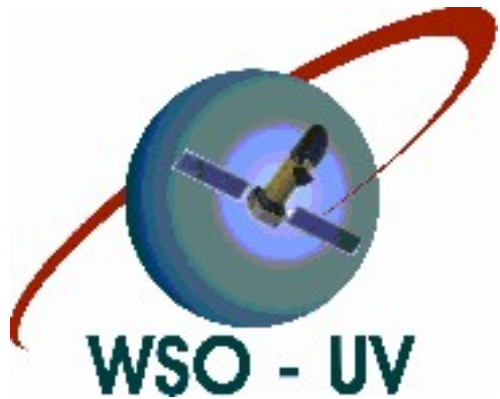

## INTRODUCTION

During the UN/ESA Workshops on Basic Space Science, the concept of a world space observatory has been recognised as an important tool to bring about the desired quantum leaps in development (United Nations, 1999).

The world space observatory embodies a twofold goal:

(a) To create opportunities for participation, at the frontiers of science, on a sustainable basis and at the national level, by all countries in the world without the need for excessive investment (Bahcall et al., 1989; Labeyrie and Lena, 1990). In so doing, the observatory will make an important contribution to the development of an academically mature and competitive cadre in many developing countries within 5 to 10 years after inception of the project in 1996 by offering equal opportunities to astronomers all over the world;

(b) To support world wide collaboration and to ensure that the study of the mysteries of the universe from space can be maintained in a sustainable way by scientists from all countries. This will then not only maintain the curiosity-driven spirit of discovery that is an integral part of sustainable development, but also make a reality in the scientific world of the visionary principle that space is the province of all mankind (CEOS, 2002).

**THE WORLD SPACE OBSERVATORY: FROM CONCEPT TO REALITY**

The world space observatory would consist of a satellite observatory in a context that extends beyond the normal planning of the major space agencies. The new approach incorporated into the planning and launching of the world space observatory could result in significant cost savings as well as facilitate to a considerable degree the participation in the space sciences of currently "non-space-faring" nations. It would thus contribute to vigorous space science activity in the future.

The model selected will constitute a missing element in the range of tools available at present to the astrophysical community for the exploration of the universe, extending from the near solar environment to the far distant phases of evolution, when the basic building blocks of human life were being created. Even though the world space observatory has, in the first instance, been defined in the context of the ultraviolet domain, the extension of the concept to other areas that require operations based in space would be an obvious bonus that could have a major impact on the way in which research in basic space science is conducted world-wide.

The scientific needs in the ultraviolet domain have been clearly expressed by the international astrophysics community, as, for example, in the discussion at the ESA/NASA conference held in Seville, Spain, in November 1997 (Wamsteker and Gonzalez Riestra, 1997). In 1997, a working group was established that implemented the following mandate:

(a) It defined a conceptual baseline from which to determine the issues and scientific areas in which such an observatory will have a major impact (see section Scientific Objectives below);

(b) It evaluated and defined the possible applications of innovative organisational and other configurations in a world space observatory (see section Mission Concept below);

(c) It prepared for the presentation of the goals of the world space observatory as a major activity in the context of the space science with the active participation of developing countries (see section Operational Principles below).

**SCIENTIFIC OBJECTIVES**

The scientific focal points of the activities of the observatory in the ultraviolet domain are the following (Hernanz et al., 2002; Barstow et al., 2002): Planetary system science, stellar science, white dwarfs and the local interstellar medium, star formation, active galactic nuclei, structure of the universe, and collateral imaging surveys.

**MISSION CONCEPT**

The driving principles behind the design of the ultraviolet element are (ESA, 2000; JPL, 2001):

(a) Operation of a 1-7 metre-class telescope in Earth orbit with a spectroscopic (103–310 nanometers) and imaging capacity specific to the ultraviolet domain (115 - 310 nanometers);

(b) High throughput and optimised operational and orbital efficiency;

(c) Optimum benefit to be derived from the fact that ultraviolet cosmic background radiation is at a minimum around 200 nanometers;

(d) Direct access to basic space science for the international astrophysics and planetary science community; and

(e) Limitation of the technological developments needed for a prime science mission.

To reach the scientific goals and objectives of the mission, the project will be structured in an integrated manner, that is, contributions to project development will be integrated internationally on the basis of an evaluation of the capability of individual participants. This means integration of all activities on an international level-science, operations, data collection and maintenance and training - and will allow the international community as a whole to benefit directly from the innovative operational model used in the world space observatory.

**OPERATIONAL PRINCIPLES**

In accordance with the objectives described above, the following mission operations profile is under development (Rodriguez-Pascual, 2000):

(a) Application of innovative engineering and management methods in order to combine the various contributions of all nations participating as appropriate to their capability;

(b) Establishment of national scientific operations centres in all interested countries;

(c) Spacecraft operations to be carried out by an integrated network of mission operations centres in the main nations contributing to implementation of the mission in accordance with final orbit requirements; and

(d) Location of the organisational structure where maximum scientific, educational, and public participation can be assured.

In the near future this will require:

(a) Establishment of a number of scientific operations centres in countries expressing the wish to host them, independent of their direct contribution to the implementation of the project;

(b) Centralisation of a small number of mission operations centres to perform the minimal functions required to operate the mission;

(c) Integration of the work of all the centres involved. Because of the world wide distribution of the scientific operations centres, special attention will be paid to the co-ordination of their activities and to links with other satellite missions and terrestrial facilities; and

(d) Open access to data collected. To guarantee optimal use of the scientific data obtained by the mission, all data will be in the public domain. The scientific operations centres will publish their data after processing and quality control.

The concept of a world space observatory takes into account:

(a) Efficient observatory-like access to space;

(b) Allowing scientists from developing countries to participate in cutting-edge astrophysics in their own cultural environment; and

(c) Maintaining the serendipitous nature of space astrophysics and catering to the needs of ultraviolet astronomy beyond the specialised capabilities of existing and currently planned missions.

**CONCLUSION**

To date 14 National World Space Observatory Working Groups have been formed representing scientists from 17 countries. The total membership in these groups comprises 130 scientists from ESA member States and 70 scientists from non-ESA member States, including those from developing countries. The further development of the World Space Observatory mission implementation model will be reported continuously in astronomical scientific meetings and new results will be posted on respective sites on the world-wide-web (WWW, 2002).